\documentclass[onecolumn,12pt, english,draftcls]{IEEEtran}

\usepackage{times}
\usepackage{graphicx}
\usepackage{amsmath}
\usepackage{amssymb}
\usepackage{flushend}
\usepackage{algorithm}
\usepackage{algpseudocode}
\usepackage{amsfonts}

\usepackage[T1]{fontenc}
\usepackage{amsthm}
\usepackage{amssymb}
\theoremstyle{plain}

\theoremstyle{definition}

\newcommand{\jacln}{\mathrm{jacln}}

\newcommand\blfootnote[1]{%
  \begingroup
  \renewcommand\thefootnote{}\footnote{#1}%
  \addtocounter{footnote}{-1}%
  \endgroup}

\author{
  \IEEEauthorblockN{\normalsize Giuseppe Cocco\IEEEauthorrefmark{1}, Stephan Pfletschinger\IEEEauthorrefmark{2}, Monica Navarro\IEEEauthorrefmark{2}}
  \IEEEauthorblockA{ \small \\ \IEEEauthorrefmark{1}German Aerospace Center -- DLR\\
                                     \IEEEauthorrefmark{2}Centre Tecnol\`ogic de Telecomunicacions de Catalunya (CTTC)\\
                                     giuseppe.cocco@dlr.de, \{stephan.pfletschinger, monica.navarro\}@cttc.es
   }
}

\usepackage{cite}

\makeatother

\usepackage{babel}
\providecommand{\examplename}{Example}
\providecommand{\theoremname}{Theorem}
\begin{document}

\leftskip=.28in

\rightskip=.28in

\title{{Seek and Decode: Random Access with Physical-Layer Network Coding and Multiuser Detection}}

\maketitle
\vskip -4in

\begin{abstract}
We present a novel cross layer approach to random access (RA) that combines physical-layer network coding (PLNC) with multiuser detection (MUD). PLNC and MUD are applied jointly at the physical level in order to extract any linear combination of messages experiencing a collision. The set of combinations extracted from a whole frame is then processed by the receiver to recover the original packets. A simple pre-coding stage at the transmitting terminals allows the receiver to further increase system diversity. We derive an analytical bound on the system throughput and present simulation results for the decoding at the physical level as well as several performance measures at frame level in block fading channels, namely throughput, packet loss rate and energy efficiency. The results we present are promising and suggest that a cross layer approach leveraging on the joint use of PLNC and MUD can significantly improve the performance of RA systems.
\end{abstract}
\vskip -4in
\section{Introduction}\label{sec:intro}
\blfootnote{Part of the results presented in this article have been submitted to ICC 2014 \cite{cocc14_Seek} and NetCode 2014\cite{pfle14_Interference} conferences.}Random  access systems (RAS) are at the same time an opportunity and a challenge. An opportunity because they do not require (or require little) coordination among the transmitters. This, among other advantages, makes it possible to live together with large delays, typical, for instance, of satellite communication networks. However, if on the one hand the lack of coordination can be seen as an asset, on the other hand it brings about the issue of signals from different transmitters interfering at the receiver. The problem of collisions in RAS has been tackled in different ways like exploiting the difference in the power of the received signals \cite{Roberts1975_aloha_capture} or applying multiuser detection (MUD) methods as in code-division multiple access (CDMA) systems \cite{verdu_mu_detection}. Multi-packet reception, i.e., the capability for the receiver to decode more than one packet from a collision, has been and still is an active research field. In \cite{andrews2005_interf_overview} an overview of the main multiuser detection techniques is presented. The impact of multi-packets reception capability in slotted ALOHA systems has been studied in \cite{ghez_stability_SA_multipackets}. Another approach proposed in the literature consists in having each transmitter sending multiple replicas of the same packet within a frame. The receiver tries to decode the packets that do not experience collision \cite{Choudhury_83_DSA} or subtracts the decoded packets from the slots where their replicas are \cite{casini07_CRDSA_aloha}\cite{Liva_11_CRDSA}. The scheme proposed in \cite{casini07_CRDSA_aloha} has been enhanced in \cite{degaud_2009_advances_satel_ra} by inducing fluctuations in the received power in order to exploit the capture effect. Recently the possibility of decoding functions of colliding signals has been studied in \cite{zhang2006_phy_nc} and \cite{popovski06_antipackets}. In \cite{popovski06_antipackets} the linearity of error correction codes has been applied to decode the bitwise XOR of colliding signals in the two-way relay channel (TWRC) under the assumption of equal codes at both end nodes. This approach is one of the possible implementations of the wider concept of physical-layer network coding (PLNC). The performance limits for the decoding of the sum of colliding signals have been studied from an information theoretical perspective and assuming lattice codes in \cite{nazer07computation}\cite{nazer11_reliable_phy_nc}. Most part of the literature on PLNC focuses on the TWRC.
In \cite{wubben_twrc_quaternary_2010} a generalized sum-product algorithm has been proposed for PLNC in the MAC phase of the TWRC.
In \cite{pfletschinger_2011_twrc} a quaternary decoding approach for the MAC phase of the two-way relay channel has been proposed, showing that there is an advantage in obtaining the bitwise sum by combining the previously estimated individual messages rather than directly decoding the sum from the analog signal. In \cite{cocco11_mu_phy_nc_aloha} and \cite{cocco11_mu_phy_nc_MAC} PLNC has been applied for the first time to random access systems by decoding the bitwise XOR of all colliding signals within a slot and then trying to recover all transmitted packets within a frame using matrix manipulations in $\mathbb{F}_2$. In \cite{cocco12_satellite_M2M} and \cite{cocco13_ncdp} an enhanced scheme based on PLNC over extended Galois fields has been proposed, showing an increased system diversity. An information theoretical analysis of the performance of physical layer network coding in random access systems has been presented in \cite{goseling_RA_PLNC_2013} and \cite{goseling_2013}.
In the present paper we propose a random multiple access scheme for symbol-synchronous slotted ALOHA systems named \emph{Seek and Decode} (S\&D) in which each information message undergoes a precoding stage at each transmitter, than is channel encoded and finally transmitted more than once within a frame. The precoding consists in a simple multiplication by a coefficient drawn at random from an extended Galois field. The receiver tries to decode any linear combination in $\mathbb{F}_2$ from the set of colliding bursts within each slot. Once the whole frame has been processed at the physical layer, the receiver uses the set of linear combinations available to retrieve all messages transmitted within the frame by using matrix manipulation techniques over the same extended Galois field as in the pre-coding stage. The use of an extended Galois field in the pre-coding stage increases system diversity. At the physical layer the receiver employs a hybrid between a PLNC decoder and a MUD. Two different MUD schemes are considered in combination with PLNC. One is a joint decoder (JD), in which all signals are decoded together \footnote{This differs significantly from a parallel interference cancellation scheme (PIC), since in this last one several decoders are employed in parallel estimating a different message each, while in a JD just one decoder is used, which decodes jointly all of the messages.}. The other MUD technique we consider in combination with PLNC is successive interference cancellation (SIC). We derive an upper bound on the throughput at the system level and present numerical results for the number of innovative messages decoded within a slot as well as sum rate, packet loss rate and energy efficiency at frame level.
The rest of the paper is organized as follows. In Section \ref{sec:sysmod} we introduce the system model. In section \ref{sec:proposed_scheme} the proposed approach is described while in Section \ref{sec:slot_decode} we focus on the different decoding alternatives at the physical level. In Section \ref{sec:thr_an} we derive a bound on the system throughput while Section \ref{sec:num_result} contains the numerical results. Finally, the conclusions are presented in Section \ref{sec:conclusions}.
\section{System Model}\label{sec:sysmod}
\begin{figure}[tbh]
\begin{centering}
\includegraphics[width=3.5in]{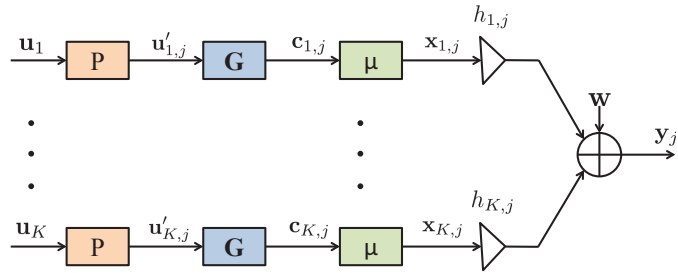}
\par\end{centering}
\caption{$K$-user multiple-access channel with block fading. All users apply
the same channel code.\label{fig:MAC}}
\end{figure}

Let us consider a random multiple access network with a population of $M$ transmitting terminals $\mathrm{T}_1, \ldots, \mathrm{T}_M$, and one receiver $\mathrm{Rx}$. In the rest of the paper we will use interchangeably the terms ``transmitting node'', ``terminal node'' and ``transmitter''. Time is divided into slots. Transmissions are organized in frames of $S$ slots each. We define a packet $\mathbf{u}$ as a block of $RN$ information bits. Each terminal generates packets according to a Poisson process of intensity $\frac{G}{M}$ packets per slot, where $G$ is the overall load offered to the network expressed in packets per slot. Each time a packet $\mathbf{u}_i=[u_{i,1},\ldots, u_{i,RN}]$ is generated at terminal $\mathrm{T}_i$, it is channel encoded  using an encoder of rate $R$ creating a codeword $\mathbf{c}_i=[c_{i,1},\ldots, c_{i,N}]$ of $N$ symbols. The same channel code is used by all transmitting nodes. The codeword $\mathbf{c}_i$ is then mapped to a binary phase-shift keying (BPSK)-modulated burst $\mathbf{x}_i$ and transmitted over the channel. We consider BPSK modulation for simplicity, but other kinds of modulations can also be used. We assume that the burst duration is approximately equal to that of a slot.
The $n$-th sample of the received signal in case of a collision of $K$ packets (also collision of \emph{size} $K$)
can then be written as
\begin{equation}
y_{n}=\sum_{k=1}^{K}h_{k}x_{k,n}+w_{n},\quad w_{n}\sim\mathcal{N}\left(0,1\right),\label{eq:yn}
\end{equation}
where the fading coefficients follow a Rayleigh distribution with
$h_{k}=\left|h_{k}^{(\mathrm{c})}\right|$ for $h_{k}^{(\mathrm{c})}\sim\mathcal{CN}\left(0,\sqrt{\mathsf{SNR}}\right)$.
The fading coefficients are known at the receiver but not at the transmitters.
 We further assume that the transmitters are synchronized such that all signals transmitted within a slot add up with symbol synchronism at the receiver. At the receiver side, $\mathrm{Rx}$ tries to decode as many linearly independent messages as possible by applying both MUD and PLNC. In order to increase system diversity at the frame level a pre-coding stage is inserted before the channel encoding at the transmitters \cite{cocco13_ncdp}. The equivalent MAC channel for a collision of size $K$ is shown on Fig. \ref{fig:MAC}, where $P$ represents the precoding, $\mathbf{G}$ is the channel coding block while $\mu$ is the mapper. We define the BPSK mapping by $x_{kn} = \mu(c_{kn})$ with $\mu(0) = -1$ and $\mu(1) = 1$.
Unlike \cite{cocco13_ncdp}, in the present work a much more efficient decoding strategy is applied by the receiver, which tries to obtain all possible linear combinations in $\mathbb{F}_2$ from the signals colliding within a slot.  Such linear combinations are then used by the receiver to recover the whole frame, treating the set of decoded linear combinations as a system of equations in $\mathbb{F}_{2^{n_{bc}}}$. The details of how different combinations are extracted from the same collision are given in Section \ref{sec:slot_decode}.
\section{Random Access with PLNC and MUD}\label{sec:proposed_scheme}
In the present section we describe the proposed random access scheme named Seek and Decode (S$\&$D). The transmitter side is the same as in \cite{cocco13_ncdp}. The main innovation is in the decoding process at both slot level and frame level. We briefly recall the operations at the transmitter side presented in \cite{cocco13_ncdp} and then move to the description of the receiver side.
\subsection{Transmitter Side}
\begin{figure}[h!]
\centerline{\includegraphics[width=3.5in]{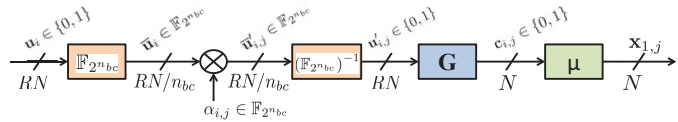}}
\caption{\footnotesize{Pre-coding, channel coding and modulation at
the transmitter side. Pre-coding consists in mapping the message to a vector in $\mathbb{F}_{2^{n_{bc}}}$, multiply each element of the vector by the same coefficient $\alpha_{i,j}$ randomly chosen in $\mathbb{F}_{2^{n_{bc}}}$ and apply an inverse mapping $\left(\mathbb{F}_{2^{n_{bc}}}\right)^{-1}$ from $\{0,1\}$ to $\mathbb{F}_{2^{n_{bc}}}$. The sub index $j$ indicates the slot within a frame in which the replica of message $\mathbf{u}_i$ is transmitted. A different coefficient $\alpha_{i,j}$ is used for each replica.}} \label{fig:tx_side}
\end{figure}

Each message is transmitted more than once within a frame, i.e., several replicas of the same message (bursts) are transmitted. Assume that node $i$ has a message $\mathbf{u}_i$ to deliver to $\mathrm{Rx}$
during a given frame, i.e., node $\mathrm{T}_i$ is an \emph{active terminal} in that frame. Before each transmission, node $i$ pre-encodes
$\mathbf{u}_i$ as depicted in Fig.~\ref{fig:tx_side}.
The message to be transmitted is divided into
 groups of $n_{bc}$ bits each. Each group of bits is mapped to a symbol in $\mathbb{F}_{2^{n_{bc}}}$ and then multiplied by a coefficient $\alpha_{i,j}\in \mathbb{F}_{2^{n_{bc}}}$. The coefficient $\alpha_{i,j}$, $j\in\{1,\ldots,S\}$ is chosen at random in each time slot $j$ while it is fixed for all symbols within a message. Note that the pre-coding does not have any impact on the decoding process at the physical layer and requires little increase in complexity with respect to a traditional scheme. The multiplication of $\mathbf{u}_i$ by $\alpha_{i,j}$  increases diversity at the frame level and does not modify the number of information bits transmitted. After the multiplication, the message is
channel-encoded, a header is attached and the modulation takes
place.
The header, which univocally determines the pre-coding coefficients, can be generated using a pseudonoise sequence generator such as the ones used in CDMA.
In practice the coefficients $\alpha_{i,j}$ can be generated using a pseudo-random number generator. In a given frame the active node chooses a different seed for the generator and uses as many outputs as the number of replicas to be transmitted. Each seed is associated to a certain header, which is detected by the receiver using the cross-correlation properties of the header\footnote{Other physical layer signatures can also be used by the nodes to allow $Rx$ to identify the transmitters. This is a subject which has been extensively studied in literature and further discussion on this is out of the scope of the present work.}. The same header is used within a given frame by an active node. In this way the receiver can detect which slots a certain node is transmitting in, and derive the coefficients used in the different replicas from the header. The header is also used to perform the channel estimation of each of the transmitters.
A more detailed analysis of the issues related to header detection and channel estimation can be found in \cite{cocco13_ncdp} and \cite{cocco12_practical_phy_nc}.
\subsection{Receiver Side}
In the literature related to random access, when a receiver receives two or more interfering signal, it can either use some kind of interference cancelation or, as in physical layer network coding, try to decode a function of the colliding signals\footnote{In \cite{lulu_pnc_mac2013} a practical implementation of a system using both PNC and MUD in the multiple access channel of a WLAN has been presented. Unlike in the present work, in \cite{lulu_pnc_mac2013} only the case of two colliding signals is considered, a relaying setup is assumed and a different multi-user detector (in which joint detection is performed but not joint decoding) is adopted.}. Most of the multiuser detection techniques found in literature can be categorized as PIC or SIC. Often such methods are iterative and alternate a detection phase to an estimation phase. In the proposed scheme the receiver applies a \emph{joint} decoder which tries to recover simultaneously all messages involved in the collision. An FFT-based belief propagation decoder over the vectorial combination of all message bits, which is described in detail in \cite{a1201}, has been adopted. The decoder jointly estimates all the single messages and then calculates the bitwise XOR of any subset of the estimated messages. It is important to notice that, as shown in \cite{pfletschinger_2011_twrc}, the sum in $\mathbb{F}_{2}$ of a set of estimated messages can be correct even if the estimated messages taken individually contain errors. A cyclic redundancy check (CRC) can be used for error detection. Note that, due to the linearity of the code, the XOR of the CRCs relative to a set of messages is a valid CRC for the XOR of the messages in the set. Here we assume ideal error detection at the receiver for ease of exposition.
Given a slot with a collision of size $K$, the receiver tries to decode $K$ independent linear combinations in $\mathbb{F}_{2}$ of the colliding signals.
Note that the total number of linear combinations that the decoder can try to recover is $\sum_{i=1}^K{K \choose i}=2^{K}-1$. Assuming the receiver is able to reliably estimate the random coefficients and the identity of the transmitters in each slot \cite{cocco13_ncdp}\cite{cocco12_practical_phy_nc}, each decoded linear combination in $\mathbb{F}_{2}$ can be interpreted at the receiver, according to arithmetics of Galois fields, as an equation in $\mathbb{F}_{2^{n_{bc}}}$. Stacking together all equations, the receiver ends up with a linear system with the form
\begin{equation}
\mathbf{A}^T\mathbf{u}=\mathbf{b},
\end{equation}
where $\bf{A}$ is the coefficient matrix having $N^{tx}$ rows and a number of columns that depends on the number of combinations decoded at PHY level, $\mathbf{u}=[\mathbf{u}_1\ldots,\mathbf{u}_{N^{tx}}]^T$ is a vector containing the different information messages transmitted in the frame, $\bf{b}$ is a vector containing the messages decoded at PHY level and $^T$ is the transpose operator.
\subsection{Example}
\begin{figure}[h!]
\centerline{\includegraphics[width=3.5in]{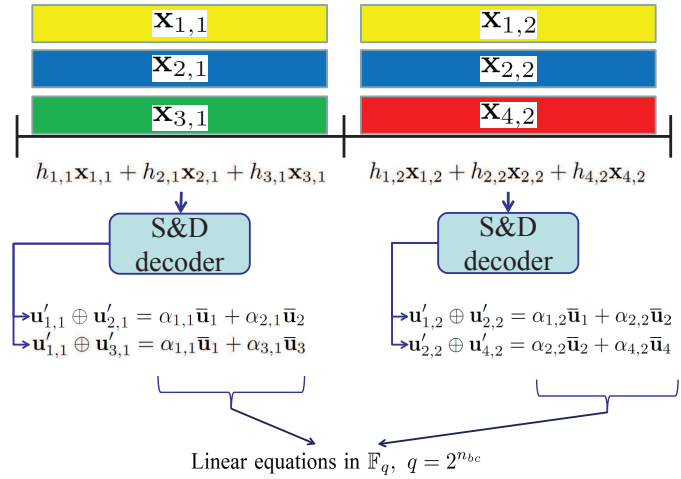}}
\caption{Example of decoding at the physical layer in S$\&$D with a two-slots frame and four active terminals. Nodes $1$ and $2$ transmit in both slots, each time choosing at random their pre-coding coefficients. Node $3$ only transmits in the first slot while node $4$ transmits only in the second. We recall that, as shown in \ref{fig:tx_side}, $\overline{\mathbf{u}}_i$ represents the mapping of the information message $\mathbf{u}_i$ form an $RN$-dimensional vector in $\mathbb{F}_2$ to an $RN/n_{bc}$-dimensional vector in $\mathbb{F}_{2^{n_{bc}}}$.} \label{fig:example}
\end{figure}

In the following we illustrate the S$\&$D scheme with a toy example. Let us consider a frame with $S=2$ slots and four active nodes.  Let us assume that nodes $1$ and $2$ transmit in both slots, each time choosing at random their pre-coding coefficients. Node $3$ only transmits in the first slot while node $4$ transmits only in the second, as illustrated in Fig. \ref{fig:example}. Let us assume that the S$\&$D decoder is able to output only two linear combinations in each of the two slots as shown in the picture. The receiver tries, then, to recover all information messages $\mathbf{u}_1,\ldots,\mathbf{u}_4$.
The decoding is possible if the coefficient matrix $\mathbf{A}$ in $\mathbb{F}_{2^{n_{bc}}}$ (shown below) has rank equal to the number of active terminals
\[ \mathbf{A}^T= \left(\begin{array}{cccc}
\alpha_{1,1} & \alpha_{2,1} & 0 & 0 \\
\alpha_{1,1} & 0 & \alpha_{3,1} & 0 \\
\alpha_{1,2} & \alpha_{2,2} & 0 & 0\\
0 & \alpha_{2,2} & 0 & \alpha_{4,2}\end{array} \right).\]
Note that coefficient $\alpha_{1,1}$ is present twice in the first column of matrix $\mathbf{A}^T$. This is because the first two rows of the matrix correspond to equations obtained from the same slot. Note also that matrix $\mathbf{A}$ is rank deficient if coefficients are chosen in $\mathbb{F}_{2}$ (i.e., all coefficients shown in the matrix above are equal to $1$), while it can be full rank if coefficients are chosen in some larger Galois field, since
the probability of obtaining a full rank matrix increases with the field size \cite{finite_fields_lidl}. Finally, we note that in the example the average number of packets decoded per slot, if $\mathbf{A}$ is full rank, is $2$.

We stress the fact that the proposed scheme does much more than simply applying a MUD, since any linear combination of the colliding signals decoded at the physical level can be exploited in the decoding phase at the frame level. In the resulting cross-layer scheme the interaction between the decoding at slot and frame level is of fundamental importance. In Section \ref{sec:slot_decode} we present different decoding approaches at PHY level.

We also note that, in principle, it would be possible to use the soft information extracted from each slot and combine it at the frame level. Although such approach could be expected to perform better than S\&D, its complexity and memory requirements would be much larger with respect to the S\&D scheme, which has the advantage of processing each slot only once and allows a lower complexity decoding at the frame level, since all operations are performed over an EGF of size $2^{n_{bc}}$, which is suited to a digital implementation.

\section{Decoding at Slot Level}\label{sec:slot_decode}
In Section \ref{sec:proposed_scheme} we described the S$\&$D scheme assuming a joint decoder is applied at the physical layer. However different kind of multi-user detector can be adapted to the S$\&$D scheme. Although some of them may lose in terms of performance with respect to the joint decoding approach, they can be attractive from a practical perspective for their lower complexity. In the present section we consider several alternative schemes while in Section \ref{sec:num_result} we numerically compare their performance in terms of the number of innovative packets decoded from a collision. Here we focus only on decoding per slot, while the performance at frame level is assessed in Section \ref{sec:num_result}.

\subsection{Separate Decoding}

The simplest approach is to decode each packet separately, considering
all other packets as interference. As for all other schemes to follow,
we utilize the CSI of all other users and the known transmit alphabet,
i.e. the BPSK1 constellation%
\footnote{A further simplification would be to consider the interference as
Gaussian noise, which would result in reduced performance and is therefore
not considered here.%
}. With this, we can write the log-likelihood value (L-value) of
user $i$ and bit position $n$ as:

\begin{equation}
L_{i,n}\triangleq\ln\frac{P\left[c_{k
i,n}=1\mid y_{n}\right]}{P\left[c_{i,n}=0\mid y_{n}\right]}=\ln\frac{P\left[x_{i,n}=1\mid y_{n}\right]}{P\left[x_{i,n}=-1\mid y_{n}\right]}.
\end{equation}

Since the received symbol $y_{n}$ depends on all symbols, we need
to marginalize over all other users' symbols. For this, we define
the sets $\mathcal{X}_{i}^{(b)}\triangleq\left\{ \mathbf{x}=\mu\left(\mathbf{d}\right):\:\mathbf{d}\in\mathbb{F}_{2}^{K},d_{i}=b\right\} $
for $b\in\mathbb{F}_{2}$, with cardinality$\left|\mathcal{X}_{i}^{(b)}\right|=2^{K-1}$
. We can think of the variable $\mathbf{d}$ as the vector of the
coded bits of all users at the same position, i.e. $\mathbf{d}_{n}=\left[c_{1,n},c_{2,n},\ldots,c_{K,n}\right]^{\mathrm{T}}$.
We obtain for the L-values

\begin{equation}
\begin{split}L_{i,n} & =\ln\frac{\sum_{\mathbf{x}\in\mathcal{X}_{i}^{(1)}}P\left[\mathbf{x\mid}y_{n}\right]}{\sum_{\mathbf{x}\in\mathcal{X}_{i}^{(0)}}P\left[\mathbf{x\mid}y_{n}\right]}=\ln\frac{\sum_{\mathbf{x}\in\mathcal{X}_{i}^{(1)}}p\left(y_{n}\mid\mathbf{x}\right)}{\sum_{\mathbf{x}\in\mathcal{X}_{i}^{(0)}}p\left(y_{n}\mid\mathbf{x}\right)}\\
 & =\ln\frac{\sum_{\mathbf{x}\in\mathcal{X}_{i}^{(1)}}\exp\left(-\left(y_{n}-\mathbf{h}^{\mathrm{T}}\mathbf{x}\right)^{2}\right)}{\sum_{\mathbf{x}\in\mathcal{X}_{i}^{(0)}}\exp\left(-\left(y_{n}-\mathbf{h}^{\mathrm{T}}\mathbf{x}\right)^{2}\right)}\\
 & =\underset{\mathbf{x}\in\mathcal{X}_{i}^{(1)}}{\jacln}\left\{ -\left(y_{n}-\mathbf{h}^{\mathrm{T}}\mathbf{x}\right)^{2}\right\} \\
 & \;\;-\underset{\mathbf{x}\in\mathcal{X}_{i}^{(0)}}{\jacln}\left\{ -\left(y_{n}-\mathbf{h}^{\mathrm{T}}\mathbf{x}\right)^{2}\right\}
\end{split}
\end{equation}
where $\jacln\left\{ x_{1},\ldots,x_{n}\right\} \triangleq\ln\sum_{j=1}^{n}\exp\left(x_{j}\right)$
denotes the Jacobian logarithm, which can be computed recursively
and for which computationally efficient approximations exist \cite{a465}.
These L-values are input to a soft-input decoder, which typically
is a Viterbi, a turbo or an LDPC decoder.

\subsection{Successive Interference Cancellation (SIC)}

A straightforward and well-known extension of basic single-user decoding
is SIC: if a packet $\mathbf{u}_{k^{*}}$ is successfully decoded,
its corresponding codeword $\mathbf{c}_{k^{*}}$ and symbol sequence
$\mathbf{x}_{k^{*}}$ are known and can be subtracted from the received
signal $\mathbf{y}_{n}$, creating a multiple-access channel with
$K-1$ users. This process can be repeated until decoding of all remaining
packets fails. To avoid unneccessary computations, we can exploit
the knowledge of the instantaneous SNRs and order the users accordingly:
let $\pi$ be a permutation of $\{1,2,\ldots,K\}$ such that
\begin{equation}
h_{\pi(1)}\geq h_{\pi(2)}\geq\cdots\geq h_{\pi(K)}
\end{equation}
then decoding starts with user $\pi(1)$. Apart from reducing computational
complexity, this ordering is also useful to reduce the probability
of undetected errors. To detect the correct decoding of a packet, in general
an additional error detection code, e.g. a CRC, has to be introduced
into each message $\mathbf{u}_{k}$. Since there is a non-zero probability
that an erroneous decoding is not detected, the number of decoding
attempts with low probability of success should be kept to a minimum.

\subsection{Seek \& Decode  with Successive Interference Cancellation (S\&D+SIC)}

For a coded slotted ALOHA system, a further decoding step after SIC
is possible. Assume that after the SIC procedure described above,
$K-K_{1}$ packets have been correctly decoded, hence leaving $K_{1}\in\{2,\ldots,K\}$
packets for which decoding failed. In this situation, the receiver
can try to decode a combined packet, which is given by the sum of
two or more of the packets that have not yet been decoded. Assume that, after SIC, users $1,2,\ldots,K_{1}$
have not been decoded. Then the receiver can try to decode a subset
of $\left\{ 1,2,\ldots,K_{1}\right\} $, e.g. given by $\mathcal{K}=\left\{ k_{1},k_{2},\ldots,k_{\ell}\right\} \subset\left\{ 1,2,\ldots,K_{1}\right\} $.
For this subset we define the sets of constellation symbols for $\ell\geq2$
as
\begin{equation}
\mathbb{X}_{\ell}^{(b)}\triangleq\left\{ \mathbf{x}=\mu\left(\mathbf{d}\right):\,\mathbf{d}\in\mathbb{F}_{2}^{\ell}\mbox{ with }\sum_{i=1}^{\ell}d_{i}=b\right\} \;,b\in\mathbb{F}_{2},\label{eq:Xl}
\end{equation}
and obtain the corresponding L-values as
\begin{equation}
L_{n}^{\mathcal{K}}=\ln\frac{\sum\limits _{\mathbf{x}\in\mathbb{X}_{\ell}^{(1)}}\exp\left(-\left(y_{n}-\left[h_{k_{1}}h_{k_{2}}\cdots h_{k_{\ell}}\right]\mathbf{x}\right)^{2}\right)}{\sum\limits _{\mathbf{x}\in\mathbb{X}_{\ell}^{(0)}}\exp\left(-\left(y_{n}-\left[h_{k_{1}}h_{k_{2}}\cdots h_{k_{\ell}}\right]\mathbf{x}\right)^{2}\right)}.\label{eq:L3}
\end{equation}

These L-values $L_{1}^{\mathcal{K}},L_{2}^{\mathcal{K}},\ldots,L_{N}^{\mathcal{K}}$
are fed to the soft-input decoder, which, if successful, finds the
corresponding codeword $\sum_{k\in\mathcal{K}}\mathbf{c}_{k}$ or
message $\sum_{k\in\mathcal{K}}\mathbf{u}_{k}$. Note that the sum
of messages or codewords is defined in the finite field $\mathbb{F}_{2}$,
which is the same as the bit-wise XOR. This concept of packet combining
is closely related to inter-flow network coding and it exploits the
linearity of the code, which can be seen by the relation
\begin{equation}
\sum_{k\in\mathcal{K}}\mathbf{c}_{k}=\sum_{k\in\mathcal{K}}\mathbf{u}_{k}\mathbf{G}.
\end{equation}
For error detection, since CRC codes are also binary linear codes,
the same CRC can be used. For $K_{1}$ undecoded packets, there exist
\[
\sum_{\ell=2}^{K_{1}}\binom{K_{1}}{\ell}=2^{K_{1}}-K_{1}-1
\]
 combinations of two or more packets, for which a decoding attempt
is possible from the L-values defined by \eqref{eq:L3}. With this
definition, note that the subsets $\mathbb{X}_{\ell}^{(b)}$ only
depend on $b$ and on the number of packets $\ell$ but not on their
indices $k_{1},\ldots,k_{\ell}$.
After successful decoding of a packet combination, a subsequent idea
is to re-apply interference cancellation with the packet combination.
This, however, is not directly possible since the combined codeword
$\sum_{k\in\mathcal{K}}\mathbf{c}_{k}$ does not correspond to any
received symbol sequence
$\mathbf{x}_{k}$
in \eqref{eq:yn}
and the sum of
codewords and symbol sequences are taken over different fields, namely
$\mathbb{F}_{2}$ and $\mathbb{R}$.
However, knowledge of a combined packet $\mathbf{c}_{\mathcal{K}}$
might still be useful for another decoding attempt: the cardinality
of the sets $\mathbb{X}_{\ell}^{(b)}$ can be reduced by a factor
of two by introducing the additional constraint of the known combined
packet. Then, the L-values can be recomputed and new decoding attempts,
including $\ell=1$ for individual packets can be undertaken. This
approach brings about a slight additional complexity due the constraint
on the decoded combination. In this case, the sets $\mathbb{X}_{\ell}^{(b)}$
will additionally depend on $n$ and hence have to computed for each
coded bit.
\subsection{Seek \& Decode with Joint Decoding (S\&D+JD)}
From \eqref{eq:yn} we can observe that, for what concerns the detection, the received samples $y_{n}$
depend on all coded bits $c_{k,n}$ at the same bit position but are
independent of bits at other positions. The optimum decoding approach
is therefore to consider the vectorial symbols $\mathbf{d}_{n}\triangleq\left[c_{1,n},c_{2,n},\ldots,c_{K,n}\right]^{\mathrm{T}}$
jointly. This can be done with a joint decoder which operates on the
vectors $\mathbf{d}_{n}$ or on an equivalent integer representation
$\bar{d}_{n}$ such that $\mathbf{d}_{n}=\mathrm{bin}(\bar{d}_{n})$.
The notation $\mathrm{bin}(b)$ denotes the binary representation
of the non-negative integer $b$. For LDPC and for convolutional codes,
such joint decoders are described in \cite{a1201,a1193}. The decoder
input is given by the probability vector
\begin{equation}
\mathbf{p}_{n}\triangleq\left[\begin{array}{c}
p_{n}(0)\\
p_{n}(1)\\
\vdots\\
p_{n}(2^{K}-1)
\end{array}\right]\in\mathbb{R}^{2^{K}},
\end{equation}
where
\begin{equation}
p_{n}(b)\triangleq P\left[\mathbf{d}=\mathrm{bin}\left(b\right)\mid y_{n}\right]\propto p\left(y_{n}\mid\mathbf{x}=\mu\left(\mathrm{bin}(b)\right)\right),
\end{equation}
 for $b=0,1,\ldots,2^{K}-1$. Let $\bar{\mathbf{x}}_{b}=\mu\left(\mathrm{bin}(b)\right)$,
then
\begin{equation}
\mathbf{p}_{n}=\alpha\left[\begin{array}{c}
\exp\left(-\left(y_{n}-\mathbf{h}^{\mathrm{T}}\bar{\mathbf{x}}_{0}\right)^{2}\right)\\
\exp\left(-\left(y_{n}-\mathbf{h}^{\mathrm{T}}\bar{\mathbf{x}}_{1}\right)^{2}\right)\\
\vdots\\
\exp\left(-\left(y_{n}-\mathbf{h}^{\mathrm{T}}\bar{\mathbf{x}}_{2^{K}-1}\right)^{2}\right)
\end{array}\right],
\end{equation}
where $\alpha$ is a scaling factor which is irrelevant for the decoding
algorithm.
The decoder output is an estimate of all messages (or equivalenty
of all codewords),
\begin{equation}
\hat{\mathbf{U}}=\left[\begin{array}{c}
\hat{\mathbf{u}}_{1}\\
\hat{\mathbf{u}}_{2}\\
\vdots\\
\hat{\mathbf{u}}_{K}
\end{array}\right].
\end{equation}
Making use of an error detecting code, the receiver checks all possible
packet combinations, i.e. all $2^{K}-1$ non-empty subsets of $\{1,2,\ldots,K\}$
and builds the binary matrix $\mathbf{A}_{slot}\in\mathbb{F}_{2}^{\left(2^{K}-1\right)\times K}$
which indicates the correct packet combinations in each row. From
this matrix, the number of innovative packets is calculated as its
rank.
This joint decoding approach reverses the order of the S\&D+SIC method:
while in S\&D+SIC the packet combination is determined first and then
a decoding attempt is carried out, joint decoding first tries to decode
all packets jointly and then the receiver checks which combinations
are correct.\\
In order to assess the performance of the different schemes considered, we count the number
of \emph{innovative }packets per slot. Innovative packets are either individually
decoded packets or combinations of packets which cannot be obtained
by combining other individually decoded packets. The number of innovative
packets is the same as the number of linearly independent packet combinations.
After successful decoding of individual packets or combinations, we
build a binary matrix $\mathbf{A}_{slot}$ whose rows $\mathbf{a}=\left[a_{1},a_{2},\ldots,a_{K}\right]$
indicate the user indices which are contained in successfully decoded
combinations. For instance, if the combined packet $\mathbf{c}_{1}+\mathbf{c}_{3}+\mathbf{c}_{4}$
is correctly decoded, the corresponding row is $\mathbf{a}=\left[1,0,1,1,0,0\right]$
for $K=6$. The number of innovative packets can then be calculated
as the rank of $\mathbf{A}_{slot}$ in $\mathbb{F}_{2}$ arithmetic.
For each slot the receiver derives a matrix $\mathbf{A}_{slot}$ over
$\mathbb{F}_2$
which is determined by the combinations decoded from a single collision. Combining the matrices from all the slots in the frame and using the information relative to the pre-coding coefficients, the receiver derives the coefficient matrix $\mathbf{A}$ introduced in Section \ref{sec:proposed_scheme}.
Another relevant benchmark we consider is joint decoding (JD), which consists in applying the joint decoder without PLNC. We adopt JD and SIC as benchmarks since they allow to measure the gains of the joint use of PLNC and MUD with respect to MUD only.
The main features of the methods presented in this section are summarized in Table \ref{tab:methods}.
\begin{table}[width=2.5in]\caption{Decoding strategies at physical layer.}\label{tab:methods}
 \centering
\begin{tabular}{|c|c|c|}
  \hline
  \textbf{Method} & \textbf{Description} & \textbf{Uses}  \\
     &  &  \textbf{precoding} \\\hline
  Separate dec. & joint detection, & no \\
    &separate decoding  &\\\hline
  SIC & joint detection,& no \\
  &separate decoding, &\\
  & then interference &\\
  & cancellation &\\\hline
  S\&D+SIC & as in SIC, & yes \\
  & then detect/decode &  \\
  & combinations &  \\\hline
   JD & joint detection, & no \\
   & joint decoding&\\\hline
   S\&D+JD & as in JD,  & yes \\
   & then combine&\\
   & estimated messages&\\\hline
\end{tabular}
\end{table}
The method JD in Table \ref{tab:methods} is another benchmark that uses joint decoding without PLNC.
\subsection{Complexity Considerations and Possible Combined Approaches}
An important aspect in the different decoding approaches at the physical layer is their performance-complexity tradeoff. For the basic
separate decoding scheme, complexity could be reduced by ordering
users according to their instantaneous SNR and stop decoding after
the decoding of one user has failed. This will obviously cause a slight
performance loss which depends mainly on the SNR differences and on
the applied coding scheme, basically on the packet length. The same
idea can be applied to both SIC techniques, while for S\&D+SIC, a
packet combination can be checked for linear independency \emph{before}
the decoding attempt. The complexity of S\&D+SIC in the worst case
is proportional to $2^{K}-1$ decoding attempts. The complexity of
joint decoding in the case of LDPC codes is proportional to $K\cdot2^{K}$
for belief propagation with transform-based check-node processing
\cite{a1172,a866}. This complexity can be reduced by applying joint
decoding after SIC and on the other hand by applying reduced-complexity
decoding algorithms \cite{a1117}.
\section{Throughput Analysis}\label{sec:thr_an}
In this section we derive an upper bound on the system throughput $\Phi$. $\Phi$ is defined as the average number of decoded messages per time slot and is a performance metric usually adopted for random access systems \cite{ghez_stability_SA_multipackets}. The throughput depends on the repetition strategy chosen. For mathematical tractability we assume a general repetition scheme in which each active node transmits in each slot with probability $p$, fixed for all nodes.
\subsection{Upper Bound on Decoding Probability}
In our simulation results in block fading channels we observed that the probability of correct decoding for the sum of a subset of messages with cardinality $i$ from a collision of size $K$, $0<i\leq K$, is a function of $i$ and $K$, apart from the rate $R$ and the average SNR. We define:
\begin{eqnarray}
Pr\{\text{decode sum of $i$ messages from collision of $K$}\} \notag \\
\triangleq p_{K,i}(R,SNR).
\end{eqnarray}
 In order to simplify the notation, in the following we fix the pair (SNR,$R$) and thus drop the dependence of the decoding probability on rate and SNR. Such dependence is implicitly assumed. $p_{K,i}$ can be upper bounded as follows:
\begin{equation}
p_{K,i} \leq \ddot{p}_{K,i}\leq \tilde{p}_{K,i},
\end{equation}
where
\begin{equation}
\ddot{p}_{K,i}\triangleq \max_{S_{K,i}} p_{K,i},
\end{equation}
$S_{K,i}$ being one of the ${K \choose i}$ subsets of $i$ messages, $i=1,\ldots, K$, among the $K$, while
\begin{equation}
\tilde{p}_{K}\triangleq \max_{i} \ddot{p}_{K,i}.
\end{equation}
 We found through simulations that $p_{K,i}$ is lower than or equal to the probability to decode the sum of the $i$ strongest signals among the $K$. Thus, $\tilde{p}_{K}$ is the maximum across all subset sizes $i$, $i\in\{1,\ldots,K\}$, of the probability to decode the sum of the $i$ strongest signals. $\tilde{p}_{K}$ is used in the following derivation.
By applying both PLNC and MUD, the receiver can obtain up to $\eta_K\triangleq 2^{K}-1$ different linear combinations in a slot with a collision of size $K$. At most $K$ of the decoded combinations are linearly independent. For ease of calculation we assume in the following that the decodability of a given combination is independent of the decoding of any other within the same slot\footnote{In general this is only an approximation, since giving the correct decoding of a subset of individual messages, any combination of such messages can also be decoded. However, it can happen that the single messages can not be decoded while the sum can (e.g., this is true for certain code rates and if two signals have the same channel amplitude as shown in \cite{pfletschinger_2011_twrc}.)}.
The number of combinations (linearly independent or not) decoded in a slot is a random variable. We indicate such variable with $\epsilon$.
Let us now indicate with $\epsilon_K$ the number of combinations decoded in a slot \emph{when the collision size is $K$}. $\epsilon_K$ is a Binomial random variable with parameters $\tilde{p}_{K,i}$ and $\eta_K$, i.e., $\epsilon_K\sim B(\eta_K,\tilde{p}_{K})$. The mean and variance of $\epsilon_K$ are $\eta_K \tilde{p}_{K}$ and $\eta_K \tilde{p}_{K}(1-\tilde{p}_{K}$), respectively.  The mean value of $\epsilon$, $E[\epsilon]=\overline{\epsilon}$, is then:
\begin{eqnarray}\label{eqn:mean_epsilon}
\overline{\epsilon} = \sum_{K=1}^{N^{tx}}{N^{tx}\choose K}p^K(1-p)^{N^{tx}-K}\eta_k\tilde{p}_{K},
\end{eqnarray}
while the mean squared value of $\epsilon$ is:
\begin{eqnarray}\label{eqn:var_epsilon}
E[\epsilon^2] = \sum_{K=1}^{N^{tx}}{N^{tx}\choose K}p^K(1-p)^{N^{tx}-K} \times\notag\\ \times\left[\eta_K \tilde{p}_{K}(1-\tilde{p}_{K}) +(\eta_K\tilde{p}_{K})^2\right].
\end{eqnarray}
We recall that $p$ is the probability of transmission in each slot for an active node, while $\tilde{p}_{K}$ is an upper bound on the probability to decode any linear combination from a collision of size $K$.
Finally, the variance of $\epsilon$, denoted in the following with $\sigma_{\epsilon}^2$, can be calculated using expressions (\ref{eqn:mean_epsilon}) and (\ref{eqn:var_epsilon}) as $\sigma_{\epsilon}^2=E[\epsilon^2]-\overline{\epsilon} ^2$. The total number of combinations decoded in the whole frame is a random variable given by the cardinality of the union of combinations decoded in all slots, which are i.i.d. random variables and for which we just calculated the mean and the variance. If $S$ is large enough, the sum of $S$ i.i.d. random variables can be approximated as a Gaussian variable having mean $S\overline{\epsilon} $ and variance $S\sigma_{\epsilon}$. From expressions (\ref{eqn:mean_epsilon}) and (\ref{eqn:var_epsilon}) it can be seen that the mean and the variance of $\epsilon$ depend on the number of active terminals in the frame. Thus we indicate with $\overline{\epsilon}(N^{tx})$ and $\sigma_{\epsilon}^2(N^{tx})$ the mean and the variance of $\epsilon$, respectively.
As mentioned in Section \ref{sec:sysmod} we assume Poisson arrivals with an overall offered load of $G$ packets per slot.  An upper bound on the normalized throughput can be calculated by assuming $p=1-2^{-n_{bc}}$, $2^{n_{bc}}$ being the size of the Galois field of the coefficients used in the pre-coding step, and assuming that all combinations decoded \emph{within a frame} are obtained using independently drawn coefficients for each message in each equation\footnote{Note that in practice each message has the same coefficient for all combinations within a given slot. Assuming independent coefficients in all equations leads to an upper bound to the probability of obtaining a matrix with rank equal to the number of active terminals.}.
The probability $Q(N^{tx},\delta, q)$ that an $N^{tx}\times (N^{tx}+\delta)$, $\delta\geq0$, random matrix in $GF(q)$ has rank $N^{tx}$ is given by \cite{finite_fields_lidl}\cite{lansberg_1893}:
\begin{eqnarray}\label{eqn:prob_full_rank}
Q(N^{tx},\delta, q)=1-\prod_{i=1}^{N^{tx}}\left(1-\frac{q^{i-1}}{q^{N^{tx}+\delta}}\right),
\end{eqnarray}
where, in our case, $q=2^{n_{bc}}$.
Using Eqn. (\ref{eqn:prob_full_rank}), the probability $p_{N^{tx}_+}$ to obtain a full rank coefficient matrix for a given number of active nodes is:
\begin{eqnarray}\label{eqn:num_dec_mean_GF}
p_{N^{tx}_+}=\sum_{N^{tx}=1}^{\infty}\frac{(GS)^{N^{tx}} e^{-GS}}{N^{tx}!}\sum_{m=N^{tx}}^{S(2^{N^{tx}}-1)}\frac{e^{-\frac{\left(m-\overline{\epsilon}_{fr}(m)\right)^2}{2\sigma_{\epsilon}^2(m)}}}{\sqrt{2\pi\sigma_{\epsilon}^2(m)}}\times\notag\\ \times Q(N^{tx},m-N^{tx}, q).
\end{eqnarray}
If $n$ is large, the probability $p_{N^{tx}_+}$ to obtain a full rank matrix coincides with the probability to decode a number of combinations larger than or equal to the number of active nodes within a frame, which is given by \eqref{eqn:num_dec_mean}
\begin{eqnarray}\label{eqn:num_dec_mean}
p_{N^{tx}_+}=\sum_{N^{tx}=1}^{\infty}\frac{(GS)^{N^{tx}} e^{-GS}}{N^{tx}!}\sum_{m=N^{tx}}^{S(2^{N^{tx}}-1)}\frac{e^{-\frac{\left(m-\overline{\epsilon}_{fr}(m)\right)^2}{2\sigma_{\epsilon}^2(m)}}}{\sqrt{2\pi\sigma_{\epsilon}^2(m)}}.
\end{eqnarray}
Using \eqref{eqn:num_dec_mean} we can obtain an upper bound on the normalized system throughput $\Phi_{UB}$ when a large field size is used. The expression for $\Phi_{UB}$ is given by Eqn. (\ref{eqn:norm_th}) at the top next page.
\begin{figure*}[t!]
\begin{small}
\begin{align}\label{eqn:norm_th}
\Phi_{UB}&=\frac{1}{S}\sum_{N^{tx}=1}^{\infty}N^{tx}\frac{(GS)^{N^{tx}} e^{-GS}}{N^{tx}!}\sum_{m=N^{tx}}^{S\left(2^{N^{tx}}-1\right)}\frac{e^{-\frac{\left(m-\overline{\epsilon}_{fr}(m)\right)^2}{2\sigma_{\epsilon}^2(m)}}}{\sqrt{2\pi\sigma_{\epsilon}^2(m)}}
&=G\sum_{N^{tx}=0}^{\infty}\frac{(GS)^{N^{tx}} e^{-GS}}{N^{tx}!}\sum_{m=N^{tx}}^{S\left(2^{N^{tx}}-1\right)}\frac{e^{-\frac{\left(m-\overline{\epsilon}_{fr}(m)\right)^2}{2\sigma_{\epsilon}^2(m)}}}{\sqrt{2\pi\sigma_{\epsilon}^2(m)}}.
\end{align}
\end{small}
\hrulefill
\end{figure*}
In practice, even if the rank of $\mathbf{A}$ is lower than $N^{tx}$ some of the packets can still be decoded using, for instance, Gaussian elimination or discarding a subset of the columns of $\mathbf{A}$ such that the remaining matrix is full rank. Furthermore, in order to have more control on the amount of power transmitted within a frame, transmitters may use a fixed number of repetitions rather than deciding in each slot whether to transmit or not.
\section{Numerical Results}\label{sec:num_result}
In the following we evaluate numerically the performance of the proposed scheme. First we compare the different PHY level decoding approaches presented in Section \ref{sec:slot_decode} in terms of the number of innovative packets decoded from a single slot, then we move to the comparison of sum rate, packet loss rate and energy efficiency at frame level for the S\&D scheme and several benchmark systems.
\vskip -3in
\subsection{Performance at Slot Level}
\begin{figure}[tbh]
\begin{centering}
\includegraphics[width=3.5in]{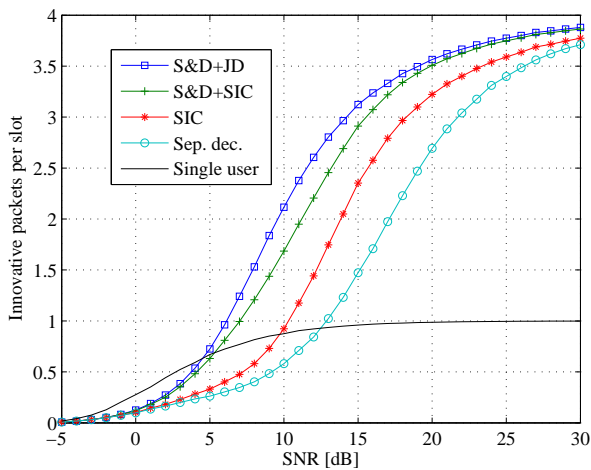}
\par\end{centering}

\caption{Innovative packets decoded per slot versus average SNR in Rayleigh fading channel for a collision of size $K=4$.\label{fig:four-user}}
\end{figure}

\begin{figure}[tbh]
\begin{centering}
\includegraphics[width=3.5in]{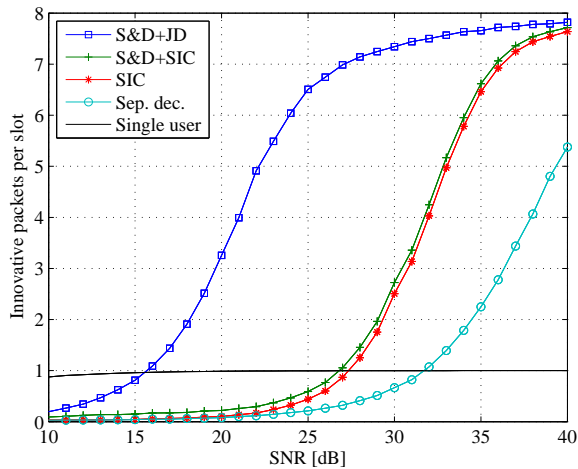}
\par\end{centering}

\caption{Innovative packets decoded per slot versus average SNR in Rayleigh fading channel for a collision of size $K=8$.\label{fig:eight-user}}
\end{figure}

We recall that innovative packets are either individually
decoded packets or combinations of packets which cannot be obtained
by combining other individually decoded packets.
Figures \ref{fig:four-user} and \ref{fig:eight-user} show the
achieved number of innovative packets per slot with the described
decoding techniques with $2$, $4$ and $8$ users, that correspond to the average rank of matrix $\mathbf{A}_{slot}$. We can see that
for all cases, S\&D+JD performs best and its advantage increases
with the number of users. For a high number of users, the advantage
of S\&D+JD to all other techniques is dramatic. On the other
hand, we point out that, unlike S\&D+JD, the S\&D+SIC scheme has the
advantage that is does not require any modification at the decoder,
since only the LLR calculation is modified with respect to a standard
receiver. We further note that the advantage of S\&D+SIC over pure
SIC decreases with the number of users. For sufficiently high SNR,
all methods achieve benefits from collided packets, which can be most
clearly seen in Fig.~\eqref{fig:four-user} for four users. At low
SNR, the average number of recovered packets per slot is close
to the single-user case while for medium to high SNR, on average
more than one packet is recovered from a single slot. For all considered cases, the number
of innovative packets tends to $K$ as the SNR grows, i.e. for high SNR nearly all
collided packets can be decoded.
\vskip -3in
\subsection{Performance at Frame Level}
\begin{figure}[h!]
  \centerline{\includegraphics[width=3.5in]{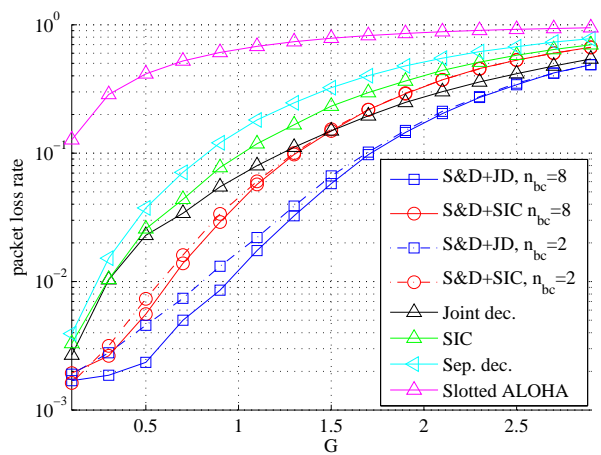}}
  \caption{Packet loss rate in  Rayleigh block fading channel, SNR=$15$ dB. The channel code is the LDPC used in WiMAX standard with parameters  N = 576, $R=1/2$, BPSK modulation. A maximum collision size of $K=7$ has been set. Collisions of higher order are discarded. The frame size $S$ has been set to $S=10$ slots. \label{fig:plr_15_dB}}
\end{figure}

\begin{figure}[h!]
  \centerline{\includegraphics[width=3.5in]{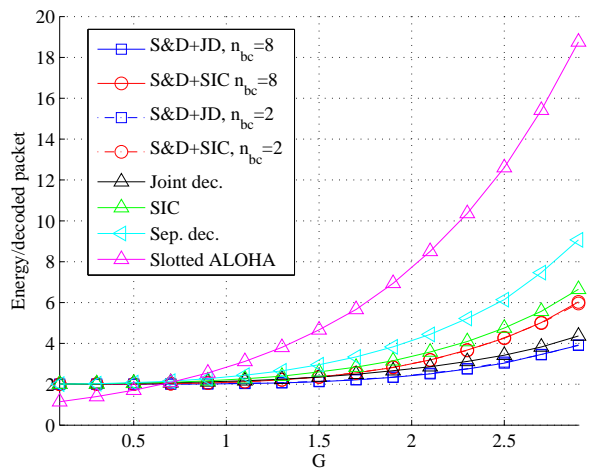}}
  \caption{Spectral efficiency plotted against energy consumption in  Rayleigh block fading channel, SNR=$15$ dB. The channel code is the WiMAX LDPC with parameters  N = 576, $R=1/2$, BPSK modulation. A maximum collision size of $K=7$ has been set. Collisions of higher order are discarded. The frame size $S$ has been set to $S=10$ slots. \label{fig:en_15_dB}}
\end{figure}

\begin{figure}[h!]
  \centerline{\includegraphics[width=3.5in]{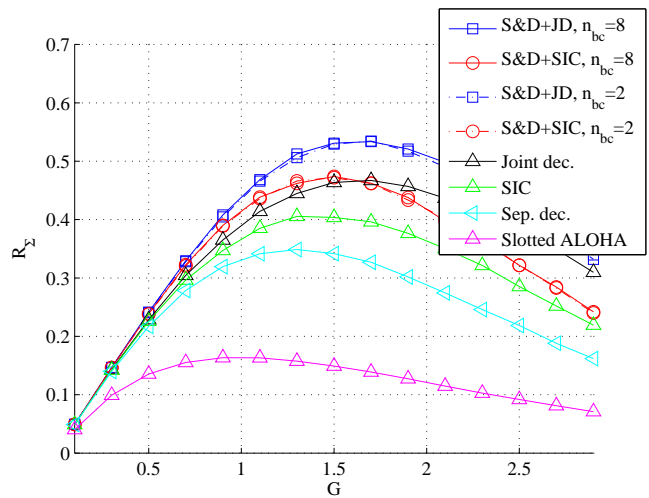}}
  \caption{Sum rate in  Rayleigh block fading channel, SNR=$10$ dB. The channel code is the WiMAX LDPC with parameters  N = 576, $R=1/2$, BPSK modulation. A maximum collision size of $K=7$ has been set. Collisions of higher order are discarded. The frame size $S$ has been set to $S=10$ slots. \label{fig:th_10_dB}}
\end{figure}

\begin{figure}[h!]
  \centerline{\includegraphics[width=3.5in]{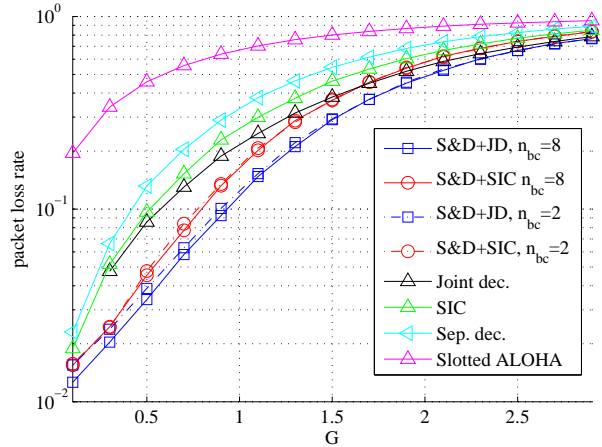}}
  \caption{Packet loss rate in  Rayleigh block fading channel, SNR=$10$ dB. The channel code is the WiMAX LDPC with parameters  N = 576, $R=1/2$, BPSK modulation. A maximum collision size of $K=7$ has been set. Collisions of higher order are discarded. The frame size $S$ has been set to $S=10$ slots. \label{fig:plr_10_dB}}
\end{figure}

\begin{figure}[h!]
  \centerline{\includegraphics[width=3.5in]{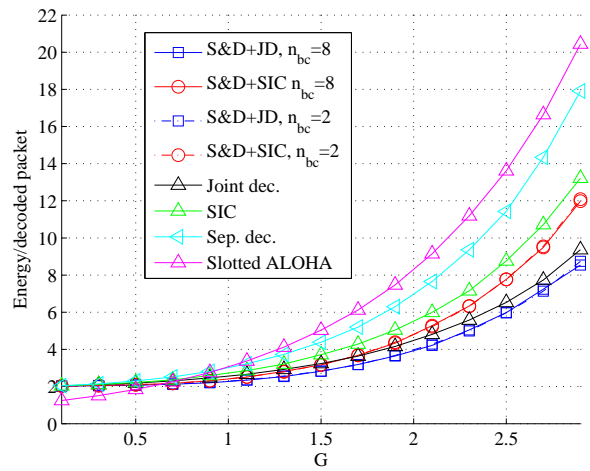}}
  \caption{Spectral efficiency plotted against energy consumption in  Rayleigh block fading channel, SNR=$10$ dB. The channel code is the WiMAX LDPC with parameters  N = 576, $R=1/2$, BPSK modulation. A maximum collision size of $K=7$ has been set. Collisions of higher order are discarded. The frame size $S$ has been set to $S=10$ slots. \label{fig:en_10_dB}}
\end{figure}

We define the sum rate $R_{\Sigma}$ as the total rate decoded within a slot averaged across the realizations. We further define the packet loss rate (PLR) as the
ratio of the number of lost packets to the total number of packets
 transmitted (not counting repetitions). The following hods:
\begin{equation}
R_{\Sigma}=RG(1-PLR),
\end{equation}
Note that $G$ is independent from the number of times
a message is repeated within a frame. Note also that the sum rate is related to the throughput as defined in Section \ref{sec:thr_an} as $R_{\Sigma}=R\Phi$.
Since the interaction between the frame and the PHY levels are of fundamental importance in the proposed scheme, in the simulations the whole decoding process has been implemented. The actual decoded combinations at the physical level have been used as input to the decoder at the frame level. As suggested in Section \ref{sec:thr_an}, if $rank(\mathbf{A})<N^{tx}$, i.e. not all messages can be decoded in a frame, the receiver applies Gaussian elimination on $\mathbf{A}$ in order to extract as many packets as possible.
\begin{figure}[h!]
  \centerline{\includegraphics[width=3.5in]{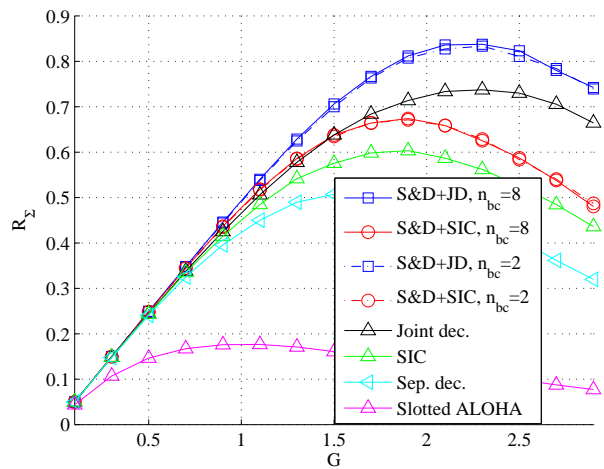}}
  \caption{Sum rate in  Rayleigh block fading channel, SNR=$15$ dB. The channel code is the WiMAX LDPC with parameters  N = 576, $R=1/2$, BPSK modulation. A maximum collision size of $K=7$ has been set. Collisions of higher order are discarded. The frame size $S$ has been set to $S=10$ slots. \label{fig:th_15_dB}}
\end{figure}
In figures \ref{fig:th_15_dB}, \ref{fig:plr_15_dB} and \ref{fig:en_15_dB} $R_{\Sigma}$, PLR and the energy efficiency are plotted against the network load $G$, respectively. The energy efficiency is defined as the ratio of the number of repetitions (which is proportional to the total amount of energy used to transmit a packet) to the number of decoded packets (not counting repetitions). Two repetitions have been considered for all schemes. A Rayleigh block fading channel with $15$ dB average SNR has been considered. The LDPC of WiMAX standard with parameters  $N = 576$, $R=1/2$, and BPSK modulation have been adopted. A maximum collision size of $k=7$ has been set, i.e., collisions of more than $7$ signals are discarded. The introduction of a maximum decodable collision size is justified by practical issues such as complexity and power saturation at the receiver.
In Fig. \ref{fig:th_15_dB} it can be seen how S\&D provides significant gains in terms of sum rate with respect to the schemes that apply MUD only. The use of larger field size in the precoding stage slightly increases the peak throughput and enhances the PLR performance at low network loads, as shown in figures \ref{fig:th_15_dB} and  \ref{fig:plr_15_dB}, respectively. It is interesting to note how the impact of the field size in S\&D is much limited with respect to the case in which only PLNC is used as in \cite{cocco13_ncdp}.
In figures \ref{fig:th_10_dB} to \ref{fig:en_10_dB} the sum rate, packet loss rate and energy efficiency for an average $SNR$ of $10$ dB are plotted, respectively.
The rest of parameters are the same as in Fig. \ref{fig:th_15_dB}.
 By comparing the two sets of figures it can be seen how the channel SNR impact the decoding at the physical layer, which leads to a higher sum rate and lower PLR when the SNR is higher, as expected. At both considered SNR values the JD scheme performs better than all others non-S\&D schemes and at $10$ dB closely approaches the S\&D+SIC for lower network loads, outperforming it in the region $G > 1.5$. Such good performance is due to the fact that the decoding of all messages is done jointly rather than separately as in the SIC or the separate decoding schemes. However, the introduction of PLNC significantly increases the performance of the JD scheme of up to a $13$ \% at both SNR values, as can be seen in Fig. \ref{fig:th_15_dB} and Fig. \ref{fig:th_10_dB}.
In all figures the slotted ALOHA scheme is also shown as a benchmark. In slotted ALOHA all terminals transmit only one replica of their message, while in all others schemes two replicas are used, i.e., twice the energy is used. In order to compare the energy efficiency of the different schemes, in Fig. \ref{fig:en_15_dB} and Fig. \ref{fig:en_10_dB} we show the average energy consumption per decoded message plotted against the load $G$. Slotted ALOHA shows a more efficient energy use at low network loads up to about $0.7$. This is due to the lower frequency of collisions and the fact that each terminals transmits half of the power used in the other schemes. However, for $G>0.7$ the other schemes, and most of all $S\&D+JD$, perform significantly better than ALOHA in terms of energy efficiency, confirming the effectiveness of the proposed approach in situations characterized by frequent collisions.
The combined decoder is capable of extracting much more information from collisions than each of the two techniques taken individually, as can be seen by comparing the results presented here with those in \cite{cocco13_ncdp}.
As a final remark, we point out that the present scheme can be used on top of other diversity schemes for ALOHA such as the one presented in \cite{Liva_11_CRDSA}. The proposed scheme would allow either to increase the throughput for a given frame size or to reduce the frame size while guaranteeing the same throughput.
\section{Conclusions}\label{sec:conclusions}
We proposed a novel cross-layer approach to random multiple access systems that uses a hybrid PLNC-MUD decoder at the physical layer and a frame level decoder based on matrix manipulation over extended Galois fields. Each node transmits several channel-coded replicas of the same message within a frame after a pre-multiplication by a  random coefficient in $\mathbb{F}_{2^{n_{bc}}}$. At the PHY layer the receiver decodes as many linear combination in $\mathbb{F}_{2}$ as possible of the signals colliding in each slot. In the second decoding stage, which is carried out at frame level, the set of combinations is seen by the receiver as a single system of equations in $\mathbb{F}_{2^{n_{bc}}}$.
We presented analytical results for the throughput at system level and simulation results for throughput, packet loss rate and energy efficiency for the case of block fading channel. In the numerical results the whole decoding process from physical to frame level has been simulated.
Our results show that a significant enhancement in throughput and PLR can be achieved by combining PLNC and MUD. The combined decoder is capable of extracting much more information from collisions than each of the two techniques taken individually.
As future work we plan
to optimize the multiple access scheme taking into account
the decoder performance, which is a function of the collision
size and the specific linear combination within a collision, with the aim of maximizing the system throughput and minimizing the PLR also taking energy efficiency into account.

As a final remark, we would like to point out that evaluating the impact of the joint use of PLNC and MUD in random access systems is a challenging task and far from being concluded. The present work can be regarded as a first step towards a full exploitation of these two techniques in the random access scenario.
\vskip -10cm
\bibliographystyle{IEEEbib}
\bibliography{PLNCMUD2014Arxiv}
\flushend
\end{document}